\documentclass[
 preprint,
 amsmath,amssymb,
 aps, prl,
]{revtex4-2}

\usepackage{graphicx}
\usepackage{dcolumn}
\usepackage{bm}
\usepackage[colorlinks, 
]{hyperref}
\usepackage[mathlines]{lineno}

\begin{document}

\preprint{APS/123-QED}

\title{Formation of Multiple Counter-propagating Clusters in the Attractive Hamiltonian Mean-field Model}

\author{Danilo M. Rivera} \email{danrivera@udec.cl}
\author{Roberto E. Navarro} \email{roberto.navarro@udec.cl}
\affiliation{%
 Departamento de F\'isica, Facultad de Ciencias F\'isicas y Matem\'aticas, Universidad de Concepci\'on. Concepci\'on, Chile.
}%

\date{\today}

\begin{abstract}
Many-body long-range interacting systems can remain approximately in a quasi-stationary state far-from-thermodynamic equilibrium. These states are typically characterized by a pair of counter-propagating density clusters, or by a single non-homogeneous core-halo in the phase-space of the particles. By using particle simulations based on the Hamiltonian mean-field model, we show that this model supports stationary states with multiple cluster or particle holes in phase-space density. We also propose a mechanism based on wave-wave and wave-particle interactions that lead to the formation of these clusters, and characterize these new quasi-stationary states in terms of the initial parameters of the simulations.
\end{abstract}

\maketitle


\section{Introduction}

Systems with long-range interactions can show particular characteristics both in their evolution towards the equilibrium state and in the equilibrium state itself \citep{Levin2014,Campa2009}. 
In particular, macroscopic quantities of the system can remain approximately in a steady state but far from the expected Boltzmann-Gibbs equilibrium state for long periods of time~\citep{Pluchino2007}. This metastable or quasi-stationary state (QSS) is characterized by non-Gaussian distribution functions~\citep{Martelloni2016}, with a lifetime that diverges with the number of particles $N$ composing the system~\citep{Roupas2020, Schutz2014}.

The Hamiltonian Mean-field (HMF) model is one of the most studied examples of long-range interacting systems \citep{Santini2022}. It has been shown that QSSs in the HMF model can be classified according to the initial total energy $u_0$ and the initial degree of homogeneity $M_0$\citep{Bachelard2008}. An homogeneous QSS is typically characterized by two counter-propagating clusters in phase-space. The formation of these two counter-propagating clusters have been explained as the result of a wave-particle interaction, where a certain group of particles resonate with counter-propagating density waves \citep{Antoniazzi2007}.

Below a certain threshold for $u_0$, a non-homogeneous QSS characterized by a core-halo structure is observed \citep{Bachelard2009, debuyl2011, Pakter2011}. This threshold corresponds to a first-order phase transition between homogeneous and non-homogeneous QSSs for $M_0<0.6$. For $M_0>0.6$, the order of this transition is unclear because the time-asymptotic state of the HMF model is highly dependent of the initial conditions~\citep{Benetti2012}. On the other hand, the formation of the core-halo structure in non-homogeneous QSSs has been linked to a Landau-like damping mechanism. Under this mechanism, particles resonate with density waves, thus gaining a large amount of energy and forming a diffuse halo in phase-space. This also results in damping of the density waves, making the remaining particles to be condensed in a dense low-energy core~\citep{Pakter2011}.


In this paper, long-term particle simulations based on the HMF model are performed. It is shown that the HMF model also supports QSSs characterized by more than two counter-propagating clusters or by core-halo distributions with an irregular halo. Depending on the initial energy $u_0$ and homogeneity degree of the distribution function, an stationary cluster can also be formed in presence of counter-propagating waves.  We classify the different QSS according to their degree of homogeneity, number, and type of structures present in phase-space, and organized depending on the properties of the initial distribution function. The classification considers both known and new cases obtained as a result of molecular dynamics simulations performed for this study.

\section{\label{sec:HMF} The Hamiltonian Mean-field model}

The HMF model corresponds to a system of $N$ particles of unitary mass moving in a circle of unit radius, interacting through a potential of infinite range \citep{Antoni1995}. The Hamiltonian describing this system is
\begin{align}
    H = \sum^N_{i=1}\frac{p_i^2}{2}+\frac{1}{2N}\sum^{N}_{i,j=1}[1-\cos(\theta_i-\theta_j)]\,,\label{Hamiltonian}
\end{align}
where $|\theta_i|<\pi$ is the position of the $i$-th particle, and $p_i$ is its conjugate momentum. The potential energy ---the second term in Eq.~\eqref{Hamiltonian}---, is considered to be attractive. Thus, the Hamilton equations of motion are:
\begin{align}
  \label{eq:motion}
  \dot\theta &= p \,, & \dot{p} &= -M_x\sin\theta_i+M_y\cos\theta_i \,,
\end{align}

where $M_x = \sum_i \cos\theta_i/N$ and $M_y= \sum_i \sin
\theta_i/N$. We also define $M=\sqrt{M_x^2+M_y^2}$, usually termed as
the magnetization, which provides a measure of the degree of
homogeneity of the system. Note that the magnetization only takes values in the domain $0\leq M\leq1$, where $M=0$ corresponds to particles distributed homogeneously in space, while $M=1$ is an extreme case where all particles are concentrated in a single point in space.

The positions and momentums of the particles are randomly initialized following a water-bag phase-space distribution, defined as
\begin{align}
    f_0(\theta,p)=\frac{1}{4\theta_0 p_0}\Theta(\theta_0-|\theta|)\Theta(p_0-|p|)\,,\label{Water-Bag}
\end{align}
where $\Theta$ is the Heaviside function, and $\theta_0\geq0$ and $p_0\geq0$ are the boundary values of the initial domain of particle positions and velocities, respectively. The initial conditions can be characterized by either $\theta_0$ and $p_0$, or by the initial energy $u_0=p_0^2/6+(1-M_0^2)/2 $ and the initial magnetization  $M_0=M_x(0)=\sin(\theta_0)/\theta_0$. Here, $M_y(0)=0$ since $f_0$ is symmetric with respect to $\theta$. In the thermodynamic limit $N \rightarrow \infty$, it can be shown that $M_y(t)=0$ at all times \citep{Levin2014}.

To study the QSS of the HMF model, $816$ particle  simulations performed. We used $N=10^6$ particles, whose equations of motion Eq.~\eqref{eq:motion} are solved using a second-order symplectic integrator with a time-step $\Delta t = 0.01$ and $2^{17}$ integration steps (corresponding to a maximum time of $t_\text{max}=1310.72$). The initial conditions were chosen in the range $0 < M_0 < 1$ and  $\frac{1}{2}(1-M_0^2) < u_0 < 2$, where the lower bound in $u_0$ corresponds to the minimum value of the total energy allowed for the HMF model when initialized with a water-bag particle distribution. The upper bound was chosen arbitrarily, however for values of $u_0>2$ no different conclusions were obtained.

\subsection{Magnetization characteristics}

\begin{figure*}[ht!]
  \includegraphics[width=\textwidth]{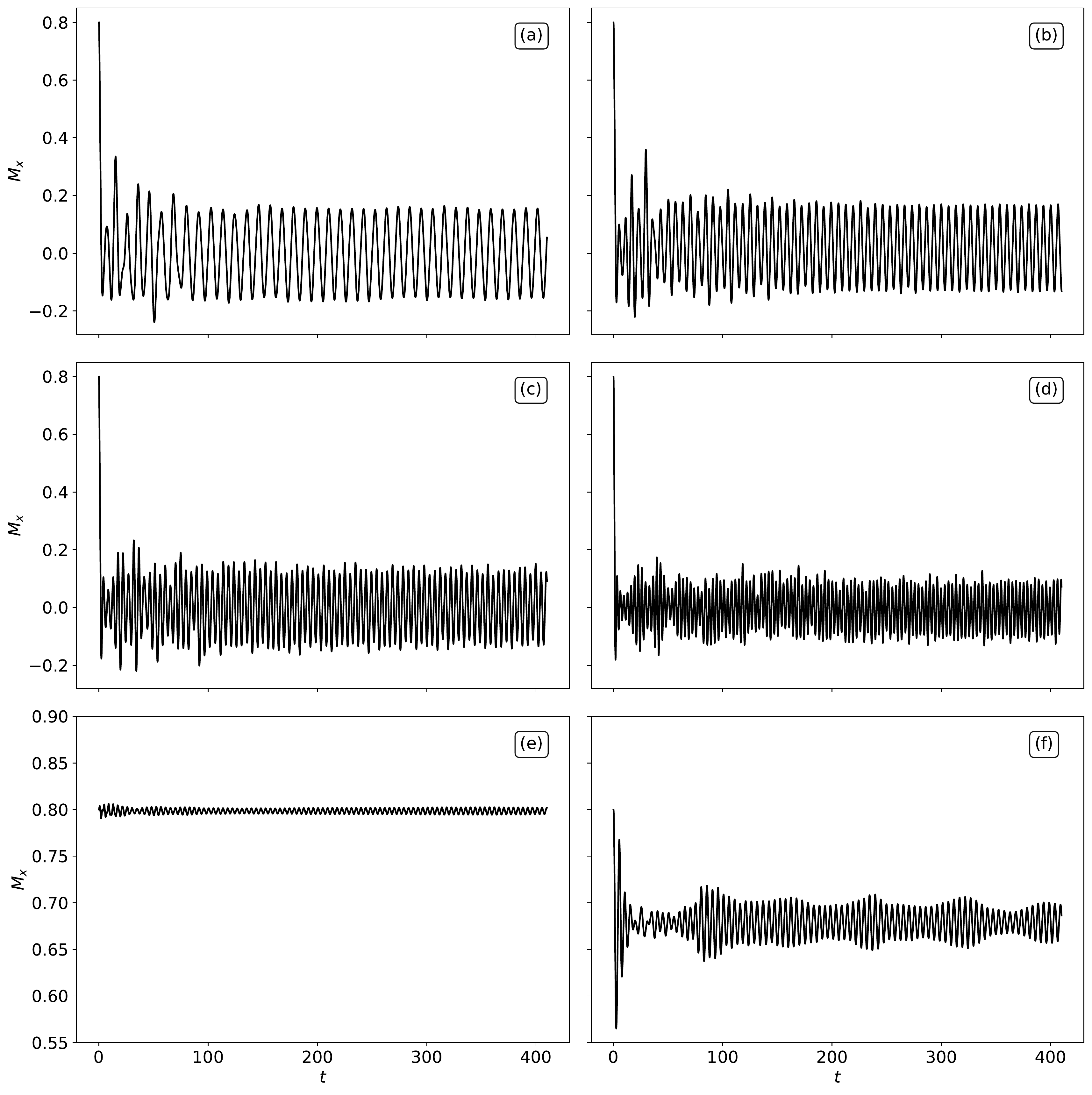}
  \caption{\label{fig:Mag} Time evolution of the magnetization $M_x(t)$ for $M_0=0.80$, $N=10^6$ particles, and (a) $u_0=0.78$, (b) $u_0=0.90$, (c) $u_0=1.10$, (d) $u_0=1.45$, (e) $u_0=0.3297$, and (f) $u_0=0.45$. Figures in the same row share axes. Notice that figures (e) and (f) use a different vertical range than figures (a)--(d).}
\end{figure*}

Figure \ref{fig:Mag} shows the time evolution of the magnetization $M_x(t)$ for six representative cases of this study, all initialized with the same magnetization $M_0=M_x(0)=0.8$. Figures \ref{fig:Mag}(a)--(d) correspond to simulations initialized with energies between $0.78\leq u_0\leq 1.45$, showing that the magnetization violently decreases and start oscillating around the time-averaged $\langle M_x\rangle=0$. Although this is expected value of the magnetization at equilibrium~\citep{Campa2009}, we will see later that the phase-space distribution function supports non-Gaussian characteristics for a long time, but eventually the distribution will relax to a Gaussian~\citep{pluchino_metastable_2004}. Although the figures show results up to $t<410$, this state remains until the end of the simulation at $t_\text{max}=1310.72$. Due to the magnetization fluctuating around $M_x=0$, these cases are usually associated to homogeneous or paramagnetic QQSs~\citep{Levin2014}. However, the term is misleading because the magnetization also exhibit strong fluctuations with amplitudes $|\delta M_x(t)|$ of the order of 0.1. These fluctuations stabilize after $t>200$, with oscillating frequency increasing with the initial energy $u_0$. Figures \ref{fig:Mag}(a) and (b) show clearly that the magnetization fluctuates at a single frequency. Meanwhile, Figs. \ref{fig:Mag}(c) and (d) show that the fluctuations in $M_x(t)$ display irregular periodicity, suggesting a bifurcation of the periodic motion~\citep{Morita2006}.

Figure \ref{fig:Mag}(e) shows the magnetization for $u_0=0.3297$ and $M_0=0.8$. In this case, $M_x(t)$ oscillates around the initial value, and with a fluctuating amplitude much lower than the discussed in Figs.~\ref{fig:Mag}(a)--(d). In fact, this case is consistent with a generalized virial condition defined by~\citet{Benetti2012}. They show that a water-bag distribution initialized with values satisfying
\begin{align}
(1-2u_0)\theta_0=\sin\theta_0 \cos\theta_0\,, \label{virial-curve}  
\end{align}
will evolve in such a way that its spatial width remains invariant, $\theta_\text{max}(t)=\theta_0$. This means that the magnetization also remains approximately invariant.

Figure \ref{fig:Mag}(d) shows the magnetization for $u_0=0.45$ and $M_0=0.8$. Here, $M_x(t)$ also violently decreases as in Figs.~\ref{fig:Mag}(a)--(d), but then oscillates around $\langle M_x\rangle \simeq 0.675$. Both cases in Figs \ref{fig:Mag}(e) and (d) are representatives of non-homogeneous or ferromagnetic QSSs~\citep{Benetti2012,Pakter2011}, i.e. whose time-averaged magnetization is $\langle M_x\rangle \neq 0.675$. The amplitude of the fluctuations is clearly modulated in the case shown in Fig. \ref{fig:Mag}(d), suggesting that at least two frequencies are at work in the evolution of the system.

\subsection{Quasi-homogeneous states}
\begin{figure*}[ht!]
  \includegraphics[width=\textwidth]{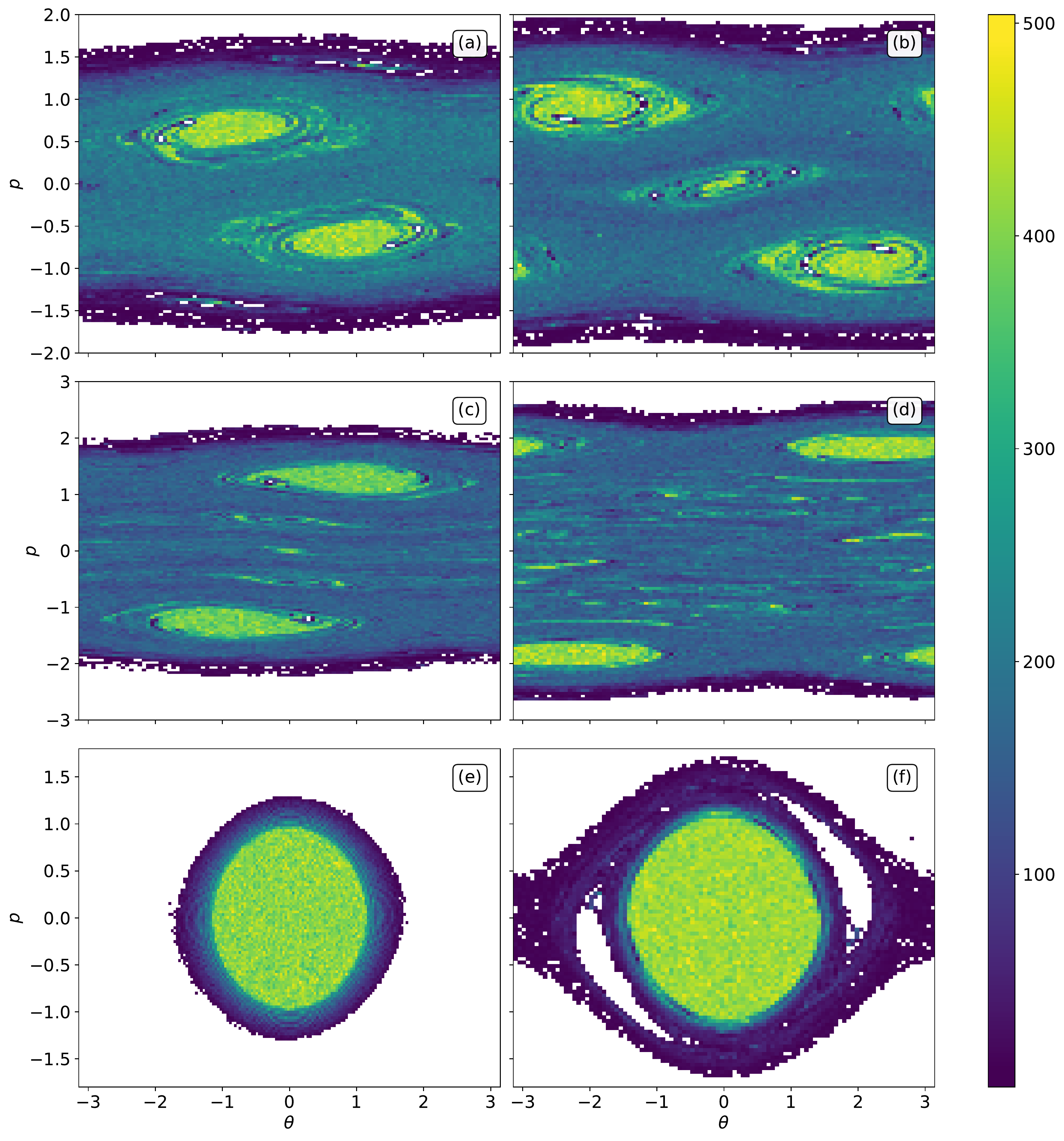}
  \caption{\label{fig:Hist2D}Phase-space density at $t=350.08$ for the same parameters as in Fig.~\ref{fig:Mag}. Figures in the same row share axes. The colorbar represents the number of particles per bin. White areas correspond to empty bins.}
\end{figure*}

Since the magnetization fluctuates in all cases shown in Fig.~\ref{fig:Mag}, even when $\langle M_x\rangle = 0$, it means that the particle density distribution is not exactly homogeneous at all times and that some structures (like clusters of particles or holes) are propagating in phase-space. This can be seen in Fig.~\ref{fig:Hist2D}, where a snapshot of the phase-space density at $t=350.08$ for the same parameters as in Fig.~\ref{fig:Mag} is shown.   

Figure~\ref{fig:Hist2D}(a) shows that the homogeneous QQS case presented in Fig.~\ref{fig:Mag}(a) is characterized by two counter-propagating clusters. The observed finite-amplitude fluctuation in the magnetization is due to the particles trying to organize themselves collectively in phase-space but failing to do so \citep{Bachelard2009}. By using phenomenological arguments, \citet{Antoniazzi2007} suggested that ---for short times---, the one-particle Hamiltonian can be understood as the Hamiltonian of one particle interacting with two counter-propagating waves with phase-speed $\Delta_p$ (although they do not show how this speed depends on the initial values of $u_0$ and $p_0$). This suggests that, depending on the initial conditions of the system, certain particles can be trapped in one of the two resonances produced by the propagation of these waves in the system, leading to the concentration of particles around the resonance regions and to the formation of these two counter-propagating clusters. Our simulations indicate that $\Delta_p$ grows mostly with the initial energy $u_0$. Due to this, as the initial energy is higher, the distance between the two counter-propagating clusters also increases. On the other hand, our simulations also indicate that the initial magnetization mostly affects how many particles compose each traveling cluster.

Since $M_y(t)=0$, then Eq.~\eqref{eq:motion} is consistent with the reflection symmetry $f(\theta,p,t)=f(-\theta,-p,t)$. This means that the counter-propagating waves must propagate at the same phase-speed and with the same amplitude. For sufficiently low energies $u_0$, then their linear superposition must form a standing wave that traps particles in the center of phase-space. This is shown in Fig.~\ref{fig:Hist2D}(b) for $u_0=0.9$, where a third cluster is centered at $\theta=0$ and $p=0$. A similar case has been reported by \citet{Yamaguchi2011,Martelloni2016}, without a proper phenomenological explanation. This third cluster appears early in the simulation, around the same time as the counter-propagating clusters. Although it deforms in time due to the traveling of the other two clusters, it conserves the area and thus the number of particles it has trapped. In the previous case discussed in Fig.~\ref{fig:Hist2D}(a), a standing wave is also present, but the counter-propagating waves travel a low enough speeds $\Delta_p$. This means that most particles are trapped by the traveling waves, with only few particles being trapped at the center of the phase-space. When $\Delta_p$ increases, the two resonance zones in phase-space are quite far apart from each other so that in the center of the phase-space there remains a non-negligible population of particles that do not interact with the two clusters, but form a third smaller stationary cluster. Thus, this case corresponds to a homogeneous QSS, but with different characteristics to the ones discussed in Fig.~\ref{fig:Hist2D}(a).

Figure~\ref{fig:Hist2D}(c) shows that, for $u_0=1.1$, a new secondary pair of counter-propagating clusters have formed around $p=\pm0.5$. For higher energies $u_0$, non-linear wave-wave coupling is expected between the traveling and stationary waves discussed in Figs.~\ref{fig:Hist2D}(b). When the system enters the quasi-stationary stage, there is a population of particles forming a streamline with speed between 0 and $\Delta_p$. Then, this streamline condenses, thus forming these secondary counter-propagating clusters. As $u_0$ increases, the density ratio between the secondary clusters and the stationary cluster also increases, meaning that the non-linear coupling becomes strong enough to carry away particles from the low-energy central cluster. This kind of QSS has been reported in \citep{Martelloni2016} and are explained through a maximum entropy scheme of the Lynden-Bell theory \citep{Lynden1967}, without linking it to a physical mechanism. 

As the energy $u_0$ increases further, and within the simulations performed, we also obtained homogeneous QSSs characterized by multiple small clusters forming spontaneously between the two counter-propagating clusters [see Fig. \ref{fig:Hist2D}(d)]. These structures are not evident during the whole quasi-stationary stage of the system: sometimes, the particles composing these small structures tend to condense making the clusters more evident, while at another stages these clusters show a dilation making their observation more complex. This may explain the irregular periodicity of the magnetization shown in Fig.~\ref{fig:Mag}(d). Moreover, these small clusters do not show to be recurrent structures, since for repeated simulations with the parameters the small structures may not emerge in phase-space density.

\subsection{Far-from-homogeneity states}
The non-homogeneous QSS characterized by $\langle M_x\rangle\neq0$. Unlike the quasi-homogeneous cases, there are no counter-propagating clusters in phase-space, but a single central structure composed of a dense core of particles surrounded by a small, sparse halo of particles, as seen in the phase-space density of Fig. \ref{fig:Hist2D}(e). \citet{Benetti2012} argued that cases initialized with initial conditions belonging to Eq.~(\ref{virial-curve}), are subject to a mean-field potential that acts on each particle. This generates only microscopic oscillations, so that parametric resonances are suppressed. This implies that the macroscopic oscillations in the system are suppressed, and magnetization fluctuations are limited as shown in Fig.~\ref{fig:Mag}(e). So, there is no physical mechanism that can cause certain particles to move to highly energetic regions, or to energetically low regions as a result of oscillation damping. The particle distribution of these QSS can be well described  by the Lyndell-Bell theory \citep{Lynden1967}, which has been shown to maximize the coarse-grained entropy characteristic of this theory \citep{Benetti2012, Levin2014}. 

For states slightly above or below the virial line Eq.~\eqref{virial-curve}, core-halo distributions can still be observed. They have been explained as a product of the Landau damping mechanism present in the system \citep{Pakter2011}. This implies that a density wave propagates in the system with which certain particles interact by gaining energies at the expense of the collective motion and reaching high energy states forming a halo. As a result of this energy transfer, the wave is damped by carrying a large number of particles to low energy states, forming the dense core.

However, as shown in Fig. \ref{fig:Hist2D}(f), the halo exhibits holes or regions free from particles in phase-space. These particle holes orbit clock wise around dense core. After averaging in phase-space, this results in the modulation of the magnetization, as shown in Fig.~\ref{fig:Mag}. The holes are not formed until the system has been in the quasi-stationary stage for some time. The presence of holes in the halo have already been reported by \citet{Benetti2012}, where they studied regular particle trajectories in the phase-space. They argue that, when the initial conditions do not coincide with the curve (\ref{virial-curve}), parametric resonances arise in the system, turning the dynamics chaotic and producing macroscopic oscillations in the system. These oscillations lead to Landau damping, which is consistent to explain the formation of the halo, but not the partial depletion of particles in the halo. The relationship between the modulation of the magnetization and these halo structures merits a detailed study.

\subsection{Classification of the phase-space distribution}

\begin{figure}[ht!]
  \includegraphics[width=\linewidth]{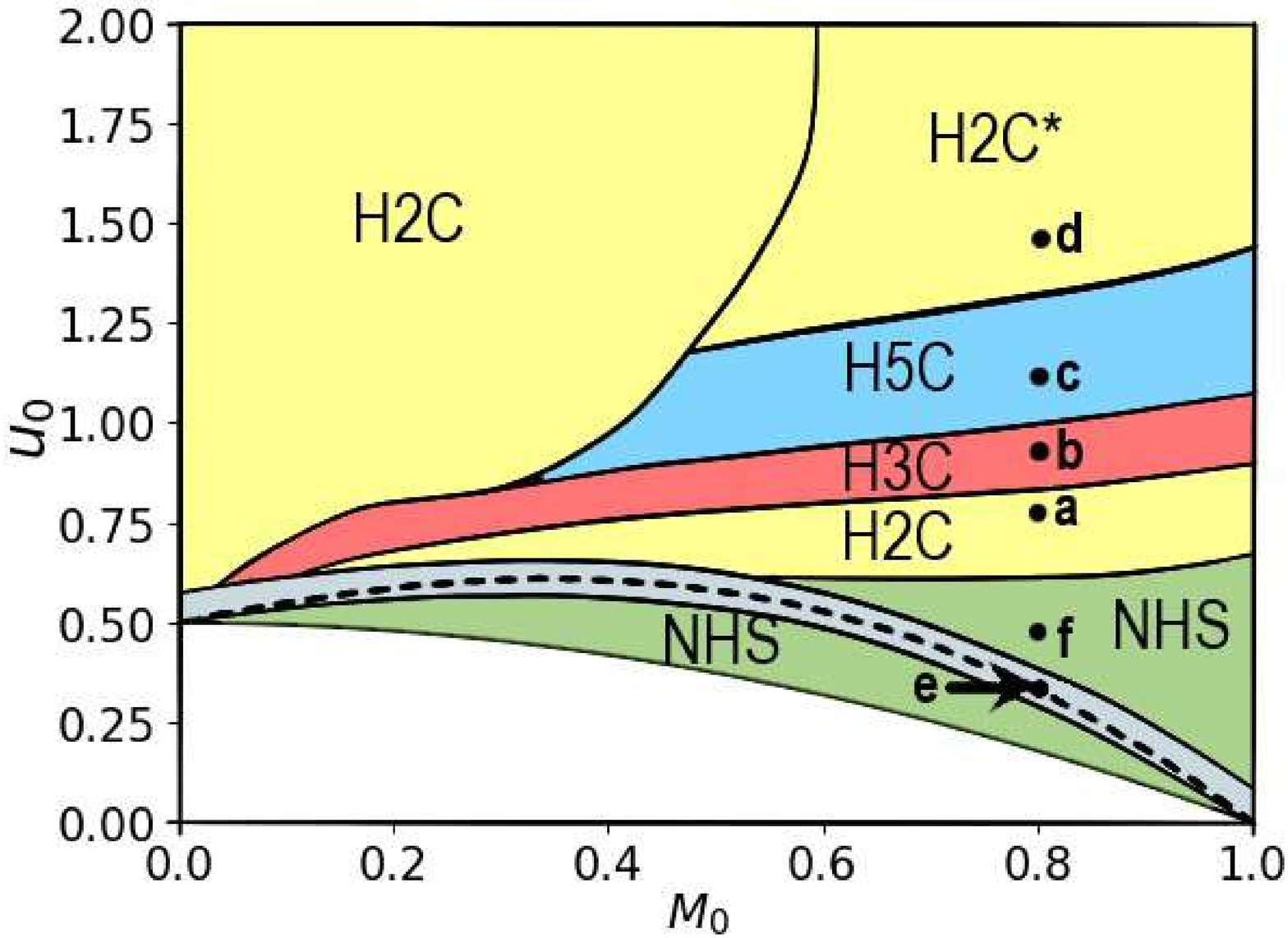}
  \caption{\label{fig:Diagram} Initial conditions space. Classification diagram of the QSS associated with different initial conditions of the HMF model. Homogeneous QSS with: (yellow--H2C) two counter-propagating clusters; (yellow--H2C*) two counter-propagating clusters and small non-recurrent clusters; (red--H3C) three clusters; (blue--H5C) five clusters. (Green--NHS) Non-homogeneous QSS with a core-halo with particle holes. The gray area  containing the dashed line corresponds to non-homogeneous QSS with a core-halo. The dashed line Eq. (\ref{virial-curve}) corresponds to non-homogeneous QSS where the magnetization of the system remains constant at all times. Points (a)-(f) are the corresponding initial states labeled in Figs. \ref{fig:Mag} and \ref{fig:Hist2D}.
}
\end{figure}

Figure~\ref{fig:Diagram} shows the phase diagram as a function of the initial parameters $M_0$ and $u_0$, where we have classified the different structures observed in the QSS phase-space distribution. Following the characteristics discussed in Figs.~\ref{fig:Mag} and \ref{fig:Hist2D}, we identify: Homogenous QSS with two clusters [H2C, as in Fig.~\ref{fig:Hist2D}(a)], three clusters [H3C, as in Fig.~\ref{fig:Hist2D}(b)], five clusters [H5C, as in Fig.~\ref{fig:Hist2D}(c)], and two clusters with multiple spontaneous low-density clusters [H2C*, as in Fig.~\ref{fig:Hist2D}(d)]. Also, non-homogeneous states (NHS) characterized by core-halo distributions with particle holes in the halo. 

The area marked as H2C* in Fig.~\ref{fig:Diagram} occur for high initial energies ($u_0>1.2$) and high initial magnetizations ($M_0> 0.5$), and we interpret it as a sub-zone of the H2C zone. There is no unique classification category for this sub-region, since it was not possible to distinguish unequivocally all the characteristics of  phase-spaces densities. However, due to the high energies, the formation of spontaneous clusters may be linked to some physical mechanism different from the one responsible for the two main clusters. 

On the other hand, Eq.~\eqref{virial-curve} determines the values (dashed lines in Fig.~\ref{fig:Diagram}) for which the magnetization does not change from the initial condition $M_0$, and where the phase-space distribution function is perfectly bounded. Due to numerical uncertainties, this curve was experimentally verified (to a small error) with the gray area in Fig.~\ref{fig:Diagram}. Note that independent of the previously commented curve (\ref{virial-curve}) is in the center of the gray zone NH which shows a concordance between the theory and the simulations.



\section{Conclusion}

Many-body long-range interacting systems can remain approximately in a quasi-stationary state far-from-thermodynamic equilibrium. These states are typically characterized by a pair of counter-propagating density clusters, or by a single non-homogeneous core-halo in the phase-space of the particles.

In this work, we analyze a set of $816$ particle simulations based on the Hamiltonian mean-field model. The initial conditions are characterized by a water-bag distribution of $N=10^6$ particles, energy $u_0$ and magnetization $M_0$. The magnetization is a measure of the homogeneity degree of the space density of particles.

We observe that the quasi-stationary states (QSS) of the HMF model are characterized by strong fluctuations in the magnetization around its time averaged value $\langle M_x(t)\rangle$, with amplitude fluctuations of the order of $0.1$. Thus, we will refer as quasi-homogeneous QSS to the cases when $\langle M_x(t)\rangle=0$ and with finite-amplitude fluctuations.  As the initial energy $u_0$ increases, the frequency of these fluctuations also increases. At some high-enough value of $u_0$, the fluctuations display irregular periodicity, suggesting a bifurcation of the periodic motion~\citep{Morita2006}. 

For quasi-homogeneous QSS, i.e. with fluctuating magnetization around $\langle M_x(t)\rangle=0$, the phase-space particle density distribution is characterized by multiple pairs of counter-propagating clusters, whose number tends to increase with the initial energy $u_0$. Since the distribution function must follow a reflection symmetry $f(x,p)=f(-x,-p)$ at all times, then each pair of counter-propagating clusters must propagate at the same speed and with the same density. This suggests the presence of counter-propagating wave densities that superpose to form a standing wave, thus trapping particles near the center $x=0$ and $p=0$ of the phase-space distribution. As the energy $u_0$ increases, the pair of counter-propagating clusters separate from each other in momentum space. Thus, through non-linear wave-wave interactions, secondary clusters are formed in phase-space. For high-enough energies, the distribution is characterized by a flat homogeneous momentum distribution for velocities bounded by the two principal counter-propagating waves, although spontaneous bumps can form through the simulation. 

For non-homogeneous QSS and for distributions far from the virial condition Eq.~\eqref{virial-curve}, the magnetization exhibits modulated finite-amplitude fluctuations around $\langle M_x(t)\rangle\neq0$. This suggests that at least two frequencies are at work through the simulation. The phase-space density in this case is characterized by a relatively dense core of particles, and a relatively tenuous halo probably formed by a Landau-like damping mechanism. In some portions of the halo, holes or regions free-from-particles can be seen, that orbit in a clock-wise direction around the core.  The holes are not formed until the system has been in the quasi-stationary stage for some time, much after the core and halo have formed.

Finally, the above description was used to characterize the QSS in terms of the initial energy $u_0$, initial magnetization $M_0$, and the number of  cluster and their characteristics in phase-space density. 

In summary, we reinterpret the notion of homogeneous QSS; provide examples where standing clusters or multiple counter-propagating clusters exist in the phase-space of particles; and provide possible mechanisms that may be responsible for the formation of these clusters. Thus, we suggest that other phase-transitions may occur, not only from a non-homogeneous to a quasi-homogeneous QSS, but also in the formation of multiple clusters that may be related to the periodicity and modulation of the density waves, as suggested by~\citep{Morita2006}.

\begin{acknowledgments}
This work has been financially supported by ANID, Chile through the projects FONDECyT No.~11180947 and 1191351 (R.E.N.). We thank Nicol Guti\'errez for useful comments.
\end{acknowledgments}


\begin{thebibliography}{18}%
\makeatletter
\providecommand \@ifxundefined [1]{%
 \@ifx{#1\undefined}
}%
\providecommand \@ifnum [1]{%
 \ifnum #1\expandafter \@firstoftwo
 \else \expandafter \@secondoftwo
 \fi
}%
\providecommand \@ifx [1]{%
 \ifx #1\expandafter \@firstoftwo
 \else \expandafter \@secondoftwo
 \fi
}%
\providecommand \natexlab [1]{#1}%
\providecommand \enquote  [1]{``#1''}%
\providecommand \bibnamefont  [1]{#1}%
\providecommand \bibfnamefont [1]{#1}%
\providecommand \citenamefont [1]{#1}%
\providecommand \href@noop [0]{\@secondoftwo}%
\providecommand \href [0]{\begingroup \@sanitize@url \@href}%
\providecommand \@href[1]{\@@startlink{#1}\@@href}%
\providecommand \@@href[1]{\endgroup#1\@@endlink}%
\providecommand \@sanitize@url [0]{\catcode `\\12\catcode `\$12\catcode
  `\&12\catcode `\#12\catcode `\^12\catcode `\_12\catcode `\%12\relax}%
\providecommand \@@startlink[1]{}%
\providecommand \@@endlink[0]{}%
\providecommand \url  [0]{\begingroup\@sanitize@url \@url }%
\providecommand \@url [1]{\endgroup\@href {#1}{\urlprefix }}%
\providecommand \urlprefix  [0]{URL }%
\providecommand \Eprint [0]{\href }%
\providecommand \doibase [0]{https://doi.org/}%
\providecommand \selectlanguage [0]{\@gobble}%
\providecommand \bibinfo  [0]{\@secondoftwo}%
\providecommand \bibfield  [0]{\@secondoftwo}%
\providecommand \translation [1]{[#1]}%
\providecommand \BibitemOpen [0]{}%
\providecommand \bibitemStop [0]{}%
\providecommand \bibitemNoStop [0]{.\EOS\space}%
\providecommand \EOS [0]{\spacefactor3000\relax}%
\providecommand \BibitemShut  [1]{\csname bibitem#1\endcsname}%
\let\auto@bib@innerbib\@empty
\bibitem [{\citenamefont {Levin}\ \emph {et~al.}(2014)\citenamefont {Levin},
  \citenamefont {Pakter}, \citenamefont {Rizzato}, \citenamefont {Teles},\ and\
  \citenamefont {Benetti}}]{Levin2014}%
  \BibitemOpen
  \bibfield  {author} {\bibinfo {author} {\bibfnamefont {Y.}~\bibnamefont
  {Levin}}, \bibinfo {author} {\bibfnamefont {R.}~\bibnamefont {Pakter}},
  \bibinfo {author} {\bibfnamefont {F.~B.}\ \bibnamefont {Rizzato}}, \bibinfo
  {author} {\bibfnamefont {T.~N.}\ \bibnamefont {Teles}},\ and\ \bibinfo
  {author} {\bibfnamefont {F.~P.}\ \bibnamefont {Benetti}},\ }\bibfield
  {title} {\bibinfo {title} {Nonequilibrium statistical mechanics of systems
  with long-range interactions},\ }\href
  {https://doi.org/https://doi.org/10.1016/j.physrep.2013.10.001} {\bibfield
  {journal} {\bibinfo  {journal} {Physics Reports}\ }\textbf {\bibinfo {volume}
  {535}},\ \bibinfo {pages} {1} (\bibinfo {year} {2014})}\BibitemShut {NoStop}%
\bibitem [{\citenamefont {Campa}\ \emph {et~al.}(2009)\citenamefont {Campa},
  \citenamefont {Dauxois},\ and\ \citenamefont {Ruffo}}]{Campa2009}%
  \BibitemOpen
  \bibfield  {author} {\bibinfo {author} {\bibfnamefont {A.}~\bibnamefont
  {Campa}}, \bibinfo {author} {\bibfnamefont {T.}~\bibnamefont {Dauxois}},\
  and\ \bibinfo {author} {\bibfnamefont {S.}~\bibnamefont {Ruffo}},\ }\bibfield
   {title} {\bibinfo {title} {Statistical mechanics and dynamics of solvable
  models with long-range interactions},\ }\href
  {https://doi.org/https://doi.org/10.1016/j.physrep.2009.07.001} {\bibfield
  {journal} {\bibinfo  {journal} {Physics Reports}\ }\textbf {\bibinfo {volume}
  {480}},\ \bibinfo {pages} {57} (\bibinfo {year} {2009})}\BibitemShut
  {NoStop}%
\bibitem [{\citenamefont {Pluchino}\ \emph {et~al.}(2007)\citenamefont
  {Pluchino}, \citenamefont {Rapisarda},\ and\ \citenamefont
  {Tsallis}}]{Pluchino2007}%
  \BibitemOpen
  \bibfield  {author} {\bibinfo {author} {\bibfnamefont {A.}~\bibnamefont
  {Pluchino}}, \bibinfo {author} {\bibfnamefont {A.}~\bibnamefont
  {Rapisarda}},\ and\ \bibinfo {author} {\bibfnamefont {C.}~\bibnamefont
  {Tsallis}},\ }\bibfield  {title} {\bibinfo {title} {Nonergodicity and
  central-limit behavior for long-range hamiltonians},\ }\href
  {https://doi.org/10.1209/0295-5075/80/26002} {\bibfield  {journal} {\bibinfo
  {journal} {Europhysics Letters ({EPL})}\ }\textbf {\bibinfo {volume} {80}},\
  \bibinfo {pages} {26002} (\bibinfo {year} {2007})}\BibitemShut {NoStop}%
\bibitem [{\citenamefont {Martelloni}\ \emph {et~al.}(2016)\citenamefont
  {Martelloni}, \citenamefont {Martelloni}, \citenamefont {de~Buyl},\ and\
  \citenamefont {Fanelli}}]{Martelloni2016}%
  \BibitemOpen
  \bibfield  {author} {\bibinfo {author} {\bibfnamefont {G.}~\bibnamefont
  {Martelloni}}, \bibinfo {author} {\bibfnamefont {G.}~\bibnamefont
  {Martelloni}}, \bibinfo {author} {\bibfnamefont {P.}~\bibnamefont
  {de~Buyl}},\ and\ \bibinfo {author} {\bibfnamefont {D.}~\bibnamefont
  {Fanelli}},\ }\bibfield  {title} {\bibinfo {title} {Generalized maximum
  entropy approach to quasistationary states in long-range systems},\ }\href
  {https://doi.org/10.1103/PhysRevE.93.022107} {\bibfield  {journal} {\bibinfo
  {journal} {Phys. Rev. E}\ }\textbf {\bibinfo {volume} {93}},\ \bibinfo
  {pages} {022107} (\bibinfo {year} {2016})}\BibitemShut {NoStop}%
\bibitem [{\citenamefont {Roupas}(2020)}]{Roupas2020}%
  \BibitemOpen
  \bibfield  {author} {\bibinfo {author} {\bibfnamefont {Z.}~\bibnamefont
  {Roupas}},\ }\bibfield  {title} {\bibinfo {title} {Statistical mechanics of
  gravitational systems with regular orbits: rigid body model of vector
  resonant relaxation},\ }\href {https://doi.org/10.1088/1751-8121/ab5f7b}
  {\bibfield  {journal} {\bibinfo  {journal} {Journal of Physics A:
  Mathematical and Theoretical}\ }\textbf {\bibinfo {volume} {53}},\ \bibinfo
  {pages} {045002} (\bibinfo {year} {2020})}\BibitemShut {NoStop}%
\bibitem [{\citenamefont {Sch\"utz}\ and\ \citenamefont
  {Morigi}(2014)}]{Schutz2014}%
  \BibitemOpen
  \bibfield  {author} {\bibinfo {author} {\bibfnamefont {S.}~\bibnamefont
  {Sch\"utz}}\ and\ \bibinfo {author} {\bibfnamefont {G.}~\bibnamefont
  {Morigi}},\ }\bibfield  {title} {\bibinfo {title} {Prethermalization of atoms
  due to photon-mediated long-range interactions},\ }\href
  {https://doi.org/10.1103/PhysRevLett.113.203002} {\bibfield  {journal}
  {\bibinfo  {journal} {Phys. Rev. Lett.}\ }\textbf {\bibinfo {volume} {113}},\
  \bibinfo {pages} {203002} (\bibinfo {year} {2014})}\BibitemShut {NoStop}%
\bibitem [{\citenamefont {Santini}\ \emph {et~al.}(2022)\citenamefont
  {Santini}, \citenamefont {Giachetti},\ and\ \citenamefont
  {Casetti}}]{Santini2022}%
  \BibitemOpen
  \bibfield  {author} {\bibinfo {author} {\bibfnamefont {A.}~\bibnamefont
  {Santini}}, \bibinfo {author} {\bibfnamefont {G.}~\bibnamefont {Giachetti}},\
  and\ \bibinfo {author} {\bibfnamefont {L.}~\bibnamefont {Casetti}},\
  }\bibfield  {title} {\bibinfo {title} {Violent relaxation in the hamiltonian
  mean field model: {II}. non-equilibrium phase diagrams},\ }\href
  {https://doi.org/10.1088/1742-5468/ac4516} {\bibfield  {journal} {\bibinfo
  {journal} {Journal of Statistical Mechanics: Theory and Experiment}\ }\textbf
  {\bibinfo {volume} {2022}},\ \bibinfo {pages} {013210} (\bibinfo {year}
  {2022})}\BibitemShut {NoStop}%
\bibitem [{\citenamefont {Bachelard}\ \emph {et~al.}(2008)\citenamefont
  {Bachelard}, \citenamefont {Chandre}, \citenamefont {Fanelli}, \citenamefont
  {Leoncini},\ and\ \citenamefont {Ruffo}}]{Bachelard2008}%
  \BibitemOpen
  \bibfield  {author} {\bibinfo {author} {\bibfnamefont {R.}~\bibnamefont
  {Bachelard}}, \bibinfo {author} {\bibfnamefont {C.}~\bibnamefont {Chandre}},
  \bibinfo {author} {\bibfnamefont {D.}~\bibnamefont {Fanelli}}, \bibinfo
  {author} {\bibfnamefont {X.}~\bibnamefont {Leoncini}},\ and\ \bibinfo
  {author} {\bibfnamefont {S.}~\bibnamefont {Ruffo}},\ }\bibfield  {title}
  {\bibinfo {title} {Abundance of regular orbits and nonequilibrium phase
  transitions in the thermodynamic limit for long-range systems},\ }\href
  {https://doi.org/10.1103/PhysRevLett.101.260603} {\bibfield  {journal}
  {\bibinfo  {journal} {Phys. Rev. Lett.}\ }\textbf {\bibinfo {volume} {101}},\
  \bibinfo {pages} {260603} (\bibinfo {year} {2008})}\BibitemShut {NoStop}%
\bibitem [{\citenamefont {Antoniazzi}\ \emph {et~al.}(2007)\citenamefont
  {Antoniazzi}, \citenamefont {Fanelli}, \citenamefont {Barr\'e}, \citenamefont
  {Chavanis}, \citenamefont {Dauxois},\ and\ \citenamefont
  {Ruffo}}]{Antoniazzi2007}%
  \BibitemOpen
  \bibfield  {author} {\bibinfo {author} {\bibfnamefont {A.}~\bibnamefont
  {Antoniazzi}}, \bibinfo {author} {\bibfnamefont {D.}~\bibnamefont {Fanelli}},
  \bibinfo {author} {\bibfnamefont {J.}~\bibnamefont {Barr\'e}}, \bibinfo
  {author} {\bibfnamefont {P.-H.}\ \bibnamefont {Chavanis}}, \bibinfo {author}
  {\bibfnamefont {T.}~\bibnamefont {Dauxois}},\ and\ \bibinfo {author}
  {\bibfnamefont {S.}~\bibnamefont {Ruffo}},\ }\bibfield  {title} {\bibinfo
  {title} {Maximum entropy principle explains quasistationary states in systems
  with long-range interactions: The example of the hamiltonian mean-field
  model},\ }\href {https://doi.org/10.1103/PhysRevE.75.011112} {\bibfield
  {journal} {\bibinfo  {journal} {Phys. Rev. E}\ }\textbf {\bibinfo {volume}
  {75}},\ \bibinfo {pages} {011112} (\bibinfo {year} {2007})}\BibitemShut
  {NoStop}%
\bibitem [{\citenamefont {Bachelard}\ \emph {et~al.}(2009)\citenamefont
  {Bachelard}, \citenamefont {Chandre}, \citenamefont {Ciani}, \citenamefont
  {Fanelli},\ and\ \citenamefont {Yamaguchi}}]{Bachelard2009}%
  \BibitemOpen
  \bibfield  {author} {\bibinfo {author} {\bibfnamefont {R.}~\bibnamefont
  {Bachelard}}, \bibinfo {author} {\bibfnamefont {C.}~\bibnamefont {Chandre}},
  \bibinfo {author} {\bibfnamefont {A.}~\bibnamefont {Ciani}}, \bibinfo
  {author} {\bibfnamefont {D.}~\bibnamefont {Fanelli}},\ and\ \bibinfo {author}
  {\bibfnamefont {Y.}~\bibnamefont {Yamaguchi}},\ }\bibfield  {title} {\bibinfo
  {title} {Analytical results on the magnetization of the hamiltonian
  mean-field model},\ }\href
  {https://doi.org/https://doi.org/10.1016/j.physleta.2009.09.037} {\bibfield
  {journal} {\bibinfo  {journal} {Physics Letters A}\ }\textbf {\bibinfo
  {volume} {373}},\ \bibinfo {pages} {4239} (\bibinfo {year}
  {2009})}\BibitemShut {NoStop}%
\bibitem [{\citenamefont {de~Buyl}\ and\ \citenamefont
  {Gaspard}(2011)}]{debuyl2011}%
  \BibitemOpen
  \bibfield  {author} {\bibinfo {author} {\bibfnamefont {P.}~\bibnamefont
  {de~Buyl}}\ and\ \bibinfo {author} {\bibfnamefont {P.}~\bibnamefont
  {Gaspard}},\ }\bibfield  {title} {\bibinfo {title} {Effectiveness of mixing
  in violent relaxation},\ }\href {https://doi.org/10.1103/PhysRevE.84.061139}
  {\bibfield  {journal} {\bibinfo  {journal} {Phys. Rev. E}\ }\textbf {\bibinfo
  {volume} {84}},\ \bibinfo {pages} {061139} (\bibinfo {year}
  {2011})}\BibitemShut {NoStop}%
\bibitem [{\citenamefont {Pakter}\ and\ \citenamefont
  {Levin}(2011)}]{Pakter2011}%
  \BibitemOpen
  \bibfield  {author} {\bibinfo {author} {\bibfnamefont {R.}~\bibnamefont
  {Pakter}}\ and\ \bibinfo {author} {\bibfnamefont {Y.}~\bibnamefont {Levin}},\
  }\bibfield  {title} {\bibinfo {title} {Core-halo distribution in the
  hamiltonian mean-field model},\ }\href
  {https://doi.org/10.1103/PhysRevLett.106.200603} {\bibfield  {journal}
  {\bibinfo  {journal} {Phys. Rev. Lett.}\ }\textbf {\bibinfo {volume} {106}},\
  \bibinfo {pages} {200603} (\bibinfo {year} {2011})}\BibitemShut {NoStop}%
\bibitem [{\citenamefont {Benetti}\ \emph {et~al.}(2012)\citenamefont
  {Benetti}, \citenamefont {Teles}, \citenamefont {Pakter},\ and\ \citenamefont
  {Levin}}]{Benetti2012}%
  \BibitemOpen
  \bibfield  {author} {\bibinfo {author} {\bibfnamefont {F.~P. d.~C.}\
  \bibnamefont {Benetti}}, \bibinfo {author} {\bibfnamefont {T.~N.}\
  \bibnamefont {Teles}}, \bibinfo {author} {\bibfnamefont {R.}~\bibnamefont
  {Pakter}},\ and\ \bibinfo {author} {\bibfnamefont {Y.}~\bibnamefont
  {Levin}},\ }\bibfield  {title} {\bibinfo {title} {Ergodicity breaking and
  parametric resonances in systems with long-range interactions},\ }\href
  {https://doi.org/10.1103/PhysRevLett.108.140601} {\bibfield  {journal}
  {\bibinfo  {journal} {Phys. Rev. Lett.}\ }\textbf {\bibinfo {volume} {108}},\
  \bibinfo {pages} {140601} (\bibinfo {year} {2012})}\BibitemShut {NoStop}%
\bibitem [{\citenamefont {Antoni}\ and\ \citenamefont
  {Ruffo}(1995)}]{Antoni1995}%
  \BibitemOpen
  \bibfield  {author} {\bibinfo {author} {\bibfnamefont {M.}~\bibnamefont
  {Antoni}}\ and\ \bibinfo {author} {\bibfnamefont {S.}~\bibnamefont {Ruffo}},\
  }\bibfield  {title} {\bibinfo {title} {Clustering and relaxation in
  hamiltonian long-range dynamics},\ }\href
  {https://doi.org/10.1103/PhysRevE.52.2361} {\bibfield  {journal} {\bibinfo
  {journal} {Phys. Rev. E}\ }\textbf {\bibinfo {volume} {52}},\ \bibinfo
  {pages} {2361} (\bibinfo {year} {1995})}\BibitemShut {NoStop}%
\bibitem [{\citenamefont {Pluchino}\ \emph {et~al.}(2004)\citenamefont
  {Pluchino}, \citenamefont {Latora},\ and\ \citenamefont
  {Rapisarda}}]{pluchino_metastable_2004}%
  \BibitemOpen
  \bibfield  {author} {\bibinfo {author} {\bibfnamefont {A.}~\bibnamefont
  {Pluchino}}, \bibinfo {author} {\bibfnamefont {V.}~\bibnamefont {Latora}},\
  and\ \bibinfo {author} {\bibfnamefont {A.}~\bibnamefont {Rapisarda}},\
  }\bibfield  {title} {\bibinfo {title} {Metastable states, anomalous
  distributions and correlations in the {HMF} model},\ }\href
  {https://doi.org/10.1016/j.physd.2004.01.029} {\bibfield  {journal} {\bibinfo
   {journal} {Phys. D}\ }\textbf {\bibinfo {volume} {193}},\ \bibinfo {pages}
  {315} (\bibinfo {year} {2004})}\BibitemShut {NoStop}%
\bibitem [{\citenamefont {Morita}\ and\ \citenamefont
  {Kaneko}(2006)}]{Morita2006}%
  \BibitemOpen
  \bibfield  {author} {\bibinfo {author} {\bibfnamefont {H.}~\bibnamefont
  {Morita}}\ and\ \bibinfo {author} {\bibfnamefont {K.}~\bibnamefont
  {Kaneko}},\ }\bibfield  {title} {\bibinfo {title} {Collective oscillation in
  a hamiltonian system},\ }\href
  {https://doi.org/https://doi.org/10.1103/PhysRevLett.96.050602} {\bibfield
  {journal} {\bibinfo  {journal} {Phys. Rev. Lett.}\ }\textbf {\bibinfo
  {volume} {96}},\ \bibinfo {pages} {050602} (\bibinfo {year}
  {2006})}\BibitemShut {NoStop}%
\bibitem [{\citenamefont {Yamaguchi}(2011)}]{Yamaguchi2011}%
  \BibitemOpen
  \bibfield  {author} {\bibinfo {author} {\bibfnamefont {Y.~Y.}\ \bibnamefont
  {Yamaguchi}},\ }\bibfield  {title} {\bibinfo {title} {Construction of
  traveling clusters in the hamiltonian mean-field model by nonequilibrium
  statistical mechanics and bernstein-greene-kruskal waves},\ }\href
  {https://doi.org/10.1103/PhysRevE.84.016211} {\bibfield  {journal} {\bibinfo
  {journal} {Phys. Rev. E}\ }\textbf {\bibinfo {volume} {84}},\ \bibinfo
  {pages} {016211} (\bibinfo {year} {2011})}\BibitemShut {NoStop}%
\bibitem [{\citenamefont {Lynden-Bell}(1967)}]{Lynden1967}%
  \BibitemOpen
  \bibfield  {author} {\bibinfo {author} {\bibfnamefont {D.}~\bibnamefont
  {Lynden-Bell}},\ }\href@noop {} {\bibfield  {journal} {\bibinfo  {journal}
  {Monthly Notices of the Royal Astronomical Society}\ }\textbf {\bibinfo
  {volume} {136}},\ \bibinfo {pages} {101} (\bibinfo {year}
  {1967})}\BibitemShut {NoStop}%
\end{thebibliography}

\providecommand{\noopsort}[1]{}\providecommand{\singleletter}[1]{#1}%

\end{document}